\documentclass[preprint,showpacs,preprintnumbers,amsmath,amssymb,showkeys]{revtex4}

\usepackage[cp1251]{inputenc}

\usepackage{graphicx}
\usepackage{dcolumn}
\usepackage{bm}

\usepackage{float}
\usepackage{stmaryrd} 

\usepackage{version}

\usepackage[unicode,bookmarks,bookmarksopen,bookmarksopenlevel=0,colorlinks,%
pageanchor=true,breaklinks=true,linkcolor=red,citecolor=blue,urlcolor=red,
pdfencoding = auto, bookmarksdepth = 4]{hyperref}  

\newcommand{\const}{\textrm{const}}

\renewcommand{\theequation}{\arabic{section}.\arabic{equation}}

\begin{document}


\title{Hodograph Method and Numerical Integration  \\ of Two Hyperbolic Quasilinear  Equations. \\ Part I. The Shallow Water Equations}

\author{E.\,V.~Shiryaeva}%
 \email{shir@math.sfedu.ru}
\affiliation{%
Institute of Mathematica, Mechanics and Computer Science, \\ Southern Federal University, Russia.
}%

\author{M.\,Yu.~Zhukov}
\email{zhuk@math.sdedu.ru}
\affiliation{%
Institute of Mathematica, Mechanics and Computer Science, \\ Southern Federal University, Russia.
}%

\date{\today}

\begin{abstract}

In paper \cite{SenashovYakhno} the variant of the hodograph method based on the conservation laws for two hyperbolic quasilinear  equations of the first order is described. Using these results we propose a method which allows to reduce the Cauchy problem for the two quasilinear PDE's to the Cauchy problem for ODE's. The proposed method is actually some similar method of characteristics for a system of two hyperbolic quasilinear  equations.
The method can be used effectively in all cases, when the linear hyperbolic equation in partial derivatives of the second order with variable coefficients, resulting from the application of the hodograph method, has an explicit expression for the Riemann--Green function. One of the method's features is the possibility to construct a multi-valued solutions. In this paper we present examples of method application for solving the classical shallow water equations.

\end{abstract}




\maketitle

\section{Introduction}\label{zhshel:sec:introd}

To study the system of two quasilinear PDE's of the first order the hodograph method based on conservation laws is presented in the paper \cite{SenashovYakhno}. For the determination of the densities and the fluxes of some conservation laws a linear hyperbolic PDE of the second order is constructed. If this hyperbolic equation has  an analytical expressions for the Riemann--Green function then the solution of the original equations, as shown in \cite{SenashovYakhno}, can easily be presented in implicit analytical form.

We show that an implicit form of solution allows to construct an efficient numerical method for integration problem with initial data.
Proposed method allows also to construct multi-valued solutions of the Cauchy problem for original equations.
In particular, method can be used for solving of the shallow water equations and studying of the breaking waves. In the proposed method, a key role plays an explicit expression for the Riemann--Green function. Actually, there are quite a lot of important equations for which such a construction is feasible. These include the shallow water equations (see, \emph{i.g.} \cite{RozhdestvenskiiYanenko,Whithem}), the equations of gas dynamics for a polytropic gas \cite{RozhdestvenskiiYanenko,Whithem}, the soliton gas equations \cite{Whithem,GenaEl} (or  Born--Infeld equation), the equations of chromatography for classical isotherms \cite{RozhdestvenskiiYanenko,FerapontovTsarev_MatModel,Kuznetsov}, and the isotachophoresis and zonal electrophoresis equations \cite{BabskiiZhukovYudovichRussian,ZhukovMassTransport,ZhukovNonSteadyITP,ElaevaMM,Elaeva_ZhVM}. A large number of equations are presented, in particular, in \cite{SenashovYakhno}. Classification of equations that allow explicit relation for the Riemann--Green function is contained in the fundamental papers \cite{Copson,Courant,Ibragimov} (see also \cite{Chirkunov,Chirkunov_2}). The analysis also shows that the proposed method, in essence, is similar to the method of characteristics, applicable in the case of the two hyperbolic quasilinear  equations.

The proposed method  can be also successfully applied to verify the quality of numerical methods for solving hyperbolic equations such as finite difference methods, finite element method, finite volume method, the Riemann solver method \emph{etc}. Note that the method does not require any approximations of original problem and
the accuracy of the calculations is determined only by the precision used for the ODE's numerical methods.

The paper is organized as follows. In Secs.~\ref{zhshArXiv:sec:2}--\ref{zhshArXiv:sec:5} the slightly modified (and simplified) results of the paper \cite{SenashovYakhno} are presented. In Sec.~\ref{zhshArXiv:sec:6} the solution on the isochrone is constructed. In this section the Cauchy ODE's problem for solving original problem is also formulated. Finally, in Sec.~\ref{zhshArXiv:sec:6} we present the numerical results for the shallow water equations with periodic initial data.

\setcounter{equation}{0}

\section{Basic equations and relations}\label{zhshArXiv:sec:2}

The variant of hodograph method described in~\cite{SenashovYakhno}, with some minor modifications, allows to construct efficient numerical algorithm for solving of two hyperbolic quasilinear  equations. For completeness we repeat some results of the paper~\cite{SenashovYakhno}.

Let for a system of two hyperbolic equations, written in the Riemann invariants, we have the Cauchy problem at $t=t_0$
\begin{equation}\label{zhshArXiv:eq:2.01}
R^1_t+ \lambda^1(R^1,R^2)R^1_x=0, \quad  R^2_t+ \lambda^2(R^1,R^2)R^2_x=0,
\end{equation}
\begin{equation}\label{zhshArXiv:eq:2.02}
R^1(x,t_0)=R^1_0(x), \quad  R^2(x,t_0)=R^2_0(x),
\end{equation}
where $R^1_0(x)$, $R^2_0(x)$ are the functions determined on some interval of the axis $x$ (possibly infinite), $\lambda^1(R^1,R^2)$, $\lambda^2(R^1,R^2)$ are the charateristic directions.

We assume that for (\ref{zhshArXiv:eq:2.01}) a conservation law is valid
\begin{equation}\label{zhshArXiv:eq:2.03}
\varphi_t+\psi_x=0,
\end{equation}
where $\varphi(R^1,R^2)$ is the density, $\psi(R^1,R^2)$ is the flux.

Computing the derivatives in (\ref{zhshArXiv:eq:2.03}) and taking into account (\ref{zhshArXiv:eq:2.01}) we have
\begin{equation}\label{zhshArXiv:eq:2.04}
(\lambda^1\varphi_{R^1}-\psi_{R^1})R^1_x+(\lambda^2\varphi_{R^2}-\psi_{R^2})R^2_x=0.
\end{equation}
Sufficient conditions for the validity of equation (\ref{zhshArXiv:eq:2.04}) has the form
\begin{equation}\label{zhshArXiv:eq:2.05}
\psi_{R^1}=\lambda^1\varphi_{R^1}, \quad \psi_{R^2}=\lambda^2\varphi_{R^2}.
\end{equation}
If the derivatives of $R^1_x$, $R^2_x$ are independent then this conditions are  necessary.

The solvability conditions of the equations (\ref{zhshArXiv:eq:2.05}) give hyperbolic linear equations for the functions $\varphi(R^1,R^2)$,  $\psi(R^1,R^2)$
\begin{equation}\label{zhshArXiv:eq:2.06}
(\lambda^1-\lambda^2)\varphi_{R^1 R^2} + \lambda^1_{R^2}\varphi_{R^1} - \lambda^2_{R^1}\varphi_{R^2}=0,
\end{equation}
\begin{equation}\label{zhshArXiv:eq:2.07}
\left(\frac{1}{\lambda^1}-\frac{1}{\lambda^2}\right)\psi_{R^1 R^2}
+\left(\frac{1}{\lambda^1}\right)_{R^2}\psi_{R^1}
-\left(\frac{1}{\lambda^2}\right)_{R^1}\psi_{R^2}=0.
\end{equation}

For the system (\ref{zhshArXiv:eq:2.06}), (\ref{zhshArXiv:eq:2.07}) we set the conditions on the characteristics
\begin{equation}\label{zhshArXiv:eq:2.08}
 (\psi - \lambda^1 \varphi)\bigr|_{{R^1=r^1}}=1,\quad
 (\psi - \lambda^2 \varphi)\bigr|_{{R^2=r^2}}=-1,
\end{equation}
\begin{equation}\label{zhshArXiv:eq:2.09}
\left(\frac{\psi}{\lambda^1} -  \varphi\right)_{{R^1=r^1}}=1, \quad
\left(\frac{\psi}{\Lambda^2} - \varphi\right)_{{R^2=r^2}}=-1,
\end{equation}
where $r^1$, $r^2$ are constants which identify the characteristics.

Note that compared to~\cite{SenashovYakhno} here in the second conditions in (\ref{zhshArXiv:eq:2.08}), (\ref{zhshArXiv:eq:2.09}) we select $(-1)$ instead $0$. It allows to simplify a final solution of the problem.

\setcounter{equation}{0}

\section{Determination of the dependence $t=t(a,b)$}\label{zhshArXiv:sec:3}

This section, almost literally,  repeats the results of the paper \cite{SenashovYakhno} for some particular case. Compared to \cite{SenashovYakhno}  more simple initial data (\ref{zhshArXiv:eq:2.02}) and modified conditions (\ref{zhshArXiv:eq:2.08}), (\ref{zhshArXiv:eq:2.09}) are selected.

The conservation law (\ref{zhshArXiv:eq:2.03}) can be written as differential forms
\begin{equation}\label{zhshArXiv:eq:3.01}
d(\psi dt - \varphi dx)=\psi_x dx \wedge dt - \varphi_t  dt \wedge dx=(\varphi_t+\psi_x) dx \wedge dt=0.
\end{equation}

\begin{figure}[H]
\centering  \includegraphics[scale=0.8]{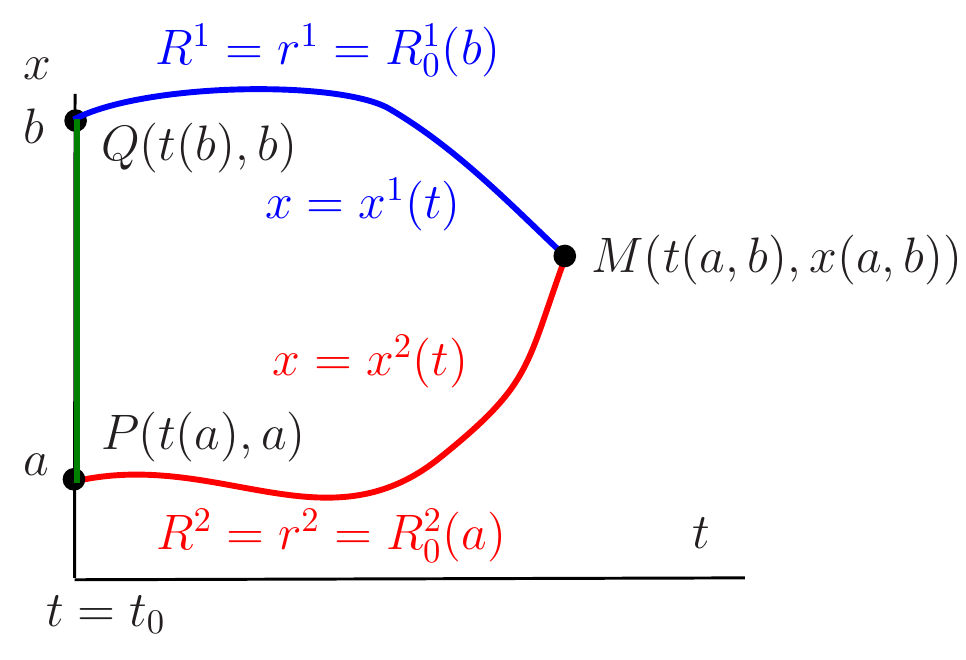}\\
  \caption{$PQM$ contour on $(x,t)$ plane}\label{zhshArXiv:fig:1.01}
\end{figure}

In the plane $(t,x)$ we choose $PQM$ contour  (see Fig.~\ref{zhshArXiv:fig:1.01}).

We assume that the lines $PM$ and $QM$ of the $PQM$ contour are the characteristics of equations (\ref{zhshArXiv:eq:2.01}), which are determined by the equations
\begin{equation}\label{zhshArXiv:eq:3.02}
QM:\quad \frac{dx^1(t)}{dt}=\lambda^1(r^1,R^2), \quad PM:\quad \frac{dx^2(t)}{dt}=\lambda^2(R^1,r^2).
\end{equation}
In other words, we have $R^1=r^1=\const$ on the $QM$ contour and  $R^2=r^2=\const$  on the $PM$ contour (see Fig.~\ref{zhshArXiv:fig:1.01}).

Path integrating the relation (\ref{zhshArXiv:eq:3.01}) over $PQM$ contour, we get
\begin{equation}\label{zhshArXiv:eq:3.03}
0=\oint\limits_{PQM}(\psi dt - \varphi dx)=
\left(\int\limits_{PQ} + \int\limits_{QM} + \int\limits_{MP}\right) (\psi dt - \varphi dx).
\end{equation}

Taking into account the relations (\ref{zhshArXiv:eq:2.08}) and (\ref{zhshArXiv:eq:3.02}) one can easily calculate integrals over $QM$ and  $MP$ contours
\begin{equation}\label{zhshArXiv:eq:3.04}
\int\limits_{QM}(\psi dt - \varphi dx)=\int\limits_{R^1=r^1}(\psi - \lambda^1 \varphi)\,dt=t-t(b),
\end{equation}
\begin{equation*}
\int\limits_{MP}(\psi dt - \varphi dx)=\int\limits_{R^2=r^2}(\psi - \lambda^2 \varphi)\,dt=t-t(a).
\end{equation*}
Using (\ref{zhshArXiv:eq:3.02}) we obtain
\begin{equation}\label{zhshArXiv:eq:3.05}
2t=t(a)+t(b)-
\int\limits_{PQ}(\psi dt - \varphi dx).
\end{equation}
The special selection of $PQM$  contour means that
\begin{equation}\label{zhshArXiv:eq:3.06}
PQ:\quad t=t_0, \quad a \leqslant x \leqslant b, \quad t(a)=t(b)=t_0.
\end{equation}

Finally, we have
\begin{equation}\label{zhshArXiv:eq:3.07}
t(a,b)=t_0+\frac12\int\limits_{a}^{b}\varphi\,dx.
\end{equation}
Note that in \cite{SenashovYakhno}, the corresponding formula is otherwise. The fact is that in \cite{SenashovYakhno} the second condition (\ref{zhshArXiv:eq:2.08}) is selected in the form: $(\psi - \lambda^2 \varphi)\bigr|_{{R^2=r^2}}=0$. This leads to the disappearance of the integral over $MP$ contour and to the unbalanced relation (\ref{zhshArXiv:eq:3.04}). In mentioned case, the integral is calculated over only one characteristic line $QM$, and the second characteristic line $MP$ is ignored. Additional simplification of the relation (\ref{zhshArXiv:eq:3.07}), compared to \cite{SenashovYakhno}, is connected to the formulation of the problem. The initial data are set at $t=t_0$ (not on an arbitrary line). This allows us to choose the line $PQ$ with the help of the relations (\ref{zhshArXiv:eq:3.06}) and eliminate the function $\psi$, since $\psi dt$ on the line $PQ$ vanishes.

It is obviously, the function $\varphi(R^1,R^2)$  also depends on the parameters $r^1$, $r^2$, $a$, and $b$. The values $r^1$, $r^2$ are determined by the initial conditions (\ref{zhshArXiv:eq:2.02})
\begin{equation}\label{zhshArXiv:eq:3.08}
  r^1=R^1_0(b), \quad  r^2=R^2_0(a).
\end{equation}
It is convenient to indicate this dependence explicitly, that is, to write  $\varphi(R^1,R^2|r^1,r^2)$.

\setcounter{equation}{0}

\section{Density $\varphi(R^1,R^2|r^1,r^2)$ and the Riemann--Green function}\label{zhshArXiv:sec:4}

We show that the function $\varphi(R^1,R^2|r^1,r^2)$ coincides with the Riemann--Green function for the equations (\ref{zhshArXiv:eq:2.06}) (accurate within factor) and  satisfies to the conditions (\ref{zhshArXiv:eq:2.08}).

Let we have the  Riemann--Green function $\Phi(R^1,R^2|r^1,r^2)$ for the equation
\begin{equation}\label{zhshArXiv:eq:4.01}
\Phi_{R^1 R^2} + A(R^1,R^2)\Phi_{R^1} + B(R^1,R^2)\Phi_{R^2}=0,
\end{equation}
\begin{equation}\label{zhshArXiv:eq:4.02}
A(R^1,R^2)=\frac{\lambda^1_{R^2}}{\lambda^1-\lambda^2}, \quad
B(R^1,R^2)=-\frac{\lambda^2_{R^1}}{\lambda^1-\lambda^2}.
\end{equation}

The function $\Phi(R^1,R^2|r^1,r^2)$ of the variables $R^1$, $R^2$ satisfies to the equation (\ref{zhshArXiv:eq:4.01}), and the function $\Phi(R^1,R^2|r^1,r^2)$ of the variables $r^1$, $r^2$ is the solution of the conjugate problem
\begin{equation}\label{zhshArXiv:eq:4.03}
\Phi_{r^1 r^2} - (A(r^1,r^2)\Phi)_{r^1} - (B(r^1,r^2)\Phi)_{r^2}=0,
\end{equation}
\begin{equation}\label{zhshArXiv:eq:4.04}
(\Phi_{r^2} - A\Phi)\bigr|_{r^1=R^1}=0, \quad (\Phi_{r^1} - B\Phi)\bigr|_{r^2=R^2}=0,
\end{equation}
\begin{equation}\label{zhshArXiv:eq:4.05}
\Phi\bigr|_{r^1=R^1,r^2=R^2}=1.
\end{equation}

We choose the Riemann--Green function  accurate within factor $M(r^1,r^2)$ as a solution of equation (\ref{zhshArXiv:eq:2.06})
\begin{equation}\label{zhshArXiv:eq:4.06}
  \varphi(R^1,R^2|r^1,r^2) = M(r^1,r^2)\Phi(R^1,R^2|r^1,r^2).
\end{equation}
It is obvious that the presence of the factor $M(r^1,r^2)$ does not affect the function $\Phi(R^1,R^2|r^1,r^2)$ of the variable $R^1$, $R^2$ is a solution of equation (\ref{zhshArXiv:eq:2.06}).

We assume that the relations (\ref{zhshArXiv:eq:2.08}) are the conditions for determination of the function $\psi(R^1,R^2)$ and multiplier $M(r^1,r^2)$. Using (\ref{zhshArXiv:eq:2.08}) we get
\begin{equation}\label{zhshArXiv:eq:4.07}
   \psi(r^1,R^2)= \lambda^1(r^1,R^2)\varphi(r^1,R^2|r^1,r^2)+1,
\end{equation}
\begin{equation*}
  \psi(R^1,r^2) = \lambda^2(R^1,r^2)\varphi(R^1,r^2|r^1,r^2)-1.
\end{equation*}
Multiplier $M(r^1,r^2)$ is easily found from matching these equations at $R^1=r^1$, $R^2=r^2$ and condition (\ref{zhshArXiv:eq:4.05})
\begin{equation}\label{zhshArXiv:eq:4.08}
  M(r^1,r^2)=\frac{2}{\lambda^2(r^1,r^2)-\lambda^1(r^1,r^2)}.
\end{equation}

Finally, we have
\begin{equation}\label{zhshArXiv:eq:4.09}
 \varphi(R^1,R^2|r^1,r^2) = \frac{2}{\lambda^2(r^1,r^2)-\lambda^1(r^1,r^2)}\Phi(R^1,R^2|r^1,r^2).
\end{equation}

The formula (\ref{zhshArXiv:eq:3.07}) takes the form
\begin{equation}\label{zhshArXiv:eq:4.10}
  t(a,b)=t_0+\frac12\int\limits_{a}^{b}\varphi(R^1_0(\tau),R^2_0(\tau)|r^1(b),r^2(a))\,d\tau,
\end{equation}
where (see~(\ref{zhshArXiv:eq:3.08}))
\begin{equation}\label{zhshArXiv:eq:4.11}
  r^1=r^1(b)=R^1_0(b), \quad r^2=r^2(a)=R^2_0(a).
\end{equation}
Note that the arguments $R^1$, $R^2$ of the function $\varphi$ are replaced by $R^1_0(\tau)$, $R^2_0(\tau)$ in~integrand, since we integrate over the $PQ$ contour (see ~(\ref{zhshArXiv:eq:3.06})).

The easiest way to determine the function $\psi(R^1,R^2)$ is the integration of the equation~(\ref{zhshArXiv:eq:2.05}) taking into account conditions (\ref{zhshArXiv:eq:4.07}). For example, rewriting the relation (\ref{zhshArXiv:eq:2.05}) in the form
\begin{equation}\label{zhshArXiv:eq:4.12}
d\psi=\lambda^1\varphi_{R^1}dR^1 + \lambda^2\varphi_{R^2}dR^2,
\end{equation}
we integrate over the contour
\begin{equation}\label{zhshArXiv:eq:4.13}
\int\limits_{(r^1,r^2)}^{(R^1,R^2)}d\psi = \psi(R^1,R^2)-\psi(r^1,r^2).
\end{equation}

Similarly, one can construct the dependency $x=x(a,b)$. Referring for details to \cite{SenashovYakhno}, we just note that it is necessary to construct the Riemann--Green function for equations (\ref{zhshArXiv:eq:2.07}) taking into account the conditions (\ref{zhshArXiv:eq:2.09}). For further, any function $\psi$, obtained using equations (\ref{zhshArXiv:eq:2.06}), (\ref{zhshArXiv:eq:2.07}) and the conditions (\ref{zhshArXiv:eq:2.09}) or a specific form of $x=x(a,b)$ are not required, and therefore, their explicit relation are not written.

\setcounter{equation}{0}

\section{Implicit form of the original Cauchy problem solution}\label{zhshArXiv:sec:5}

The results presented in Sec.~\ref{zhshArXiv:sec:4} (see also \cite{SenashovYakhno}) allow to specify an implicit form of the solution for the Cauchy problem (\ref{zhshArXiv:eq:2.01}), (\ref{zhshArXiv:eq:2.02}).

Let we have dependencies
\begin{equation}\label{zhshArXiv:eq:5.01}
t=t(a,b), \quad x=x(a,b),
\end{equation}
where $t(a,b)$ is determined by the relation (\ref{zhshArXiv:eq:4.10}), and $x(a,b)$ is the known function.

The Riemann invariants $R^1$, $R^2$ in point of $M$ with coordinates $(t(a,b),x(a,b))$ (see Fig.~\ref{zhshArXiv:fig:1.01}) are determined by the relations \begin{equation}\label{zhshArXiv:eq:5.02}
R^1(x,t)=r^1(b)=R^1_0(b), \quad R^2(x,t)=r^2(a)=R^2_0(a).
\end{equation}
Hence, the formulae (\ref{zhshArXiv:eq:5.01}), (\ref{zhshArXiv:eq:5.02}) implicitly determine the solution of the problem (\ref{zhshArXiv:eq:2.01}),
(\ref{zhshArXiv:eq:2.02}).

If the explicit solution of the (\ref{zhshArXiv:eq:5.01}) is known
\begin{equation}\label{zhshArXiv:eq:5.03}
a=a(x,t), \quad b=b(x,t)
\end{equation}
then using (\ref{zhshArXiv:eq:5.02}) one can get explicit solution of the original problem
\begin{equation}\label{zhshArXiv:eq:5.04}
R^1(x,t)=r^1(b(x,t))=R^1_0(b(x,t)), \quad R^2(x,t)=r^2(a(x,t))=R^2_0(a(x,t)).
\end{equation}

The parameters $a$, $b$ can also be interpreted as some Lagrangian variables. Value $a$, $b$ identify the `particle'
on the axis $t=t_0$ that transfer along characteristics $x=x^1(t)$, $x=x^2(t)$ the values of the invariants $R^1(b,t_0)$, $R^2(a,t_0)$ at points $a$, $b$ of axis $t=t_0$ to the point $M$ with coordinates $(t,x)$.

For the future calculations we need functions $x_a(a,b)$ and $x_b(a,b)$.
Differentiating (\ref{zhshArXiv:eq:5.04}) we obtain
\begin{equation}\label{zhshArXiv:eq:5.05}
R^1_t=r^1_b b_t, \quad  R^2_t=r^2_a a_t, \quad  R^1_x=r^1_b b_x, \quad  R^2_x=r^2_a a_x.
\end{equation}
Substituting (\ref{zhshArXiv:eq:5.05}) in (\ref{zhshArXiv:eq:2.01}) we get
\begin{equation}\label{zhshArXiv:eq:5.06}
b_t+ \lambda^1(r^1,r^2)b_x=0, \quad  a_t+ \lambda^2(r^1,r^2)a_x=0.
\end{equation}
Of course, we assume that $r^1_b$, $r^2_a$ do not vanish identically.

We emphasize that $\lambda^k(r^1,r^2)$ are given functions which are determined by initial data (\ref{zhshArXiv:eq:2.02}) and the relations (\ref{zhshArXiv:eq:3.08}) or (\ref{zhshArXiv:eq:5.02}).
\begin{equation}\label{zhshArXiv:eq:5.07}
\lambda^k(r^1,r^2)=\lambda^k(R^1_0(b),R^2_0(a)))=\lambda^k(a,b).
\end{equation}
Certainly, the system (\ref{zhshArXiv:eq:5.06}) is specific for each the Cauchy problem. The values
$a$, $b$ are the Riemann invariants for the equations (\ref{zhshArXiv:eq:5.06}).

For system (\ref{zhshArXiv:eq:5.06}) one can apply the classical hodograph method (see, \emph{i.g.} \cite{RozhdestvenskiiYanenko}). Changing role of dependent and independent variables: $(x,t) \leftrightarrow (a,b)$ we get
\begin{equation}\label{zhshArXiv:eq:5.08}
x_b=\lambda^2(r^1,r^2)t_b, \quad x_a=\lambda^1(r^1,r^2)t_a.
\end{equation}

\setcounter{equation}{0}

\section{The solution on the isochrones}\label{zhshArXiv:sec:6}

In this section, we specify a simple way, from our point of view, for construction of the solution
in the form (\ref{zhshArXiv:eq:5.01})--(\ref{zhshArXiv:eq:5.04}). To do this we reduce the original problem to the Cauchy problem for ODE's.

For simplicity we assume $t_0=0$ and introduce the notation
\begin{equation}\label{zhshArXiv:eq:6.01}
\varphi(\tau|a,b)=\frac12\varphi(R^1_0(\tau),R^2_0(\tau)|r^1(b),r^2(a)).
\end{equation}
Here, the function $\varphi(R^1,R^2|r^1,r^2)$ is deremined by (\ref{zhshArXiv:eq:4.09}).

Formula (\ref{zhshArXiv:eq:4.10}) takes the form
\begin{equation}\label{zhshArXiv:eq:6.02}
t(a,b)=\int\limits_{a}^{b}\varphi(\tau|a,b)\,d\tau, \quad t_0=0.
\end{equation}

It is easy to calculate the derivatives
\begin{equation}\label{zhshArXiv:eq:6.03}
t_a(a,b)=-\varphi(a|a,b)+\int\limits_{a}^{b}\varphi_a(\tau|a,b)\,d\tau,\quad
t_b(a,b)=\varphi(b|a,b)+\int\limits_{a}^{b}\varphi_b(\tau|a,b)\,d\tau,
\end{equation}
where
\begin{equation}\label{zhshArXiv:eq:6.04}
\varphi_a(\tau|a,b)=\frac12\varphi_{r^2}(R^1_0(\tau),R^2_0(\tau)|r^1(b),r^2(a))r^2_a(a),
\end{equation}
\begin{equation}\label{zhshArXiv:eq:6.05}
\varphi_b(\tau|a,b)=\frac12\varphi_{r^1}(R^1_0(\tau),R^2_0(\tau)|r^1(b),r^2(a))r^1_b(b).
\end{equation}
Taking into account (\ref{zhshArXiv:eq:5.08}) we get the  derivatives of $x_a$ and $x_b$
\begin{equation}\label{zhshArXiv:eq:6.06}
x_b=\lambda^2(r^1,r^2)t_b, \quad x_a=\lambda^1(r^1,r^2)t_a.
\end{equation}
We fix some value $t=t_*$ which specifies the level line (isochrone) of the function $t(a,b)$
\begin{equation}\label{zhshArXiv:eq:6.07}
t_*=t(a, b).
\end{equation}
We assume that in the plane $(a,b)$ the isochrone is  parametrically defined by the equations
\begin{equation}\label{zhshArXiv:eq:6.08}
 a=a(\tau), \quad b=b(\tau),
\end{equation}
where $\tau$ is parameter.

We select the values of $a_*$, $b_*$ which indicate some point on the isochrone $t=t_*$
\begin{equation}\label{zhshArXiv:eq:6.09}
t_*=t(a_*, b_*).
\end{equation}
In practice, the values of $a_*$, $b_*$ one can find using line levels of the function $t(a,b)$ for some ranges of parameters $a$, $b$.

To determine the coordinates $X_*=x(a_*,b_*)$ corresponding to the parameter $\tau=0$ we differentiate the function $x(a,b)$, for example, with respect to  $b$. Then, we obtain the Cauchy problem
\begin{equation}\label{zhshArXiv:eq:6.10}
\frac{dY(b)}{db}=x_b(a_*,b)=\lambda^2(r^1(b),r^2(a_*))t_b(a_*,b), \quad Y(a_*)=a_*.
\end{equation}
Integrating from $a_*$ to $b_*$ we get
\begin{equation}\label{zhshArXiv:eq:6.11}
X_*=Y(b_*).
\end{equation}

Differentiating the isochrone equation (\ref{zhshArXiv:eq:6.09}) and function $x(a,b)$ with respect to $\tau$, and taking into account (\ref{zhshArXiv:eq:6.08}), we have
\begin{equation}\label{zhshArXiv:eq:6.12}
  \frac{d t(a,b)}{d\tau} \equiv t_a(a,b) \frac{da}{d\tau} +
  t_b(a,b)\frac{db}{d\tau}=0,
\end{equation}
\begin{equation}\label{zhshArXiv:eq:6.13}
  \frac{d x(a,b)}{d\tau} \equiv x_a(a,b) \frac{da}{d\tau} +
  x_b(a,b)\frac{db}{d\tau}.
\end{equation}

The relations (\ref{zhshArXiv:eq:6.12}), (\ref{zhshArXiv:eq:6.13}) and (\ref{zhshArXiv:eq:6.06}) allow to formulate the Cauchy problem for determination
of the functions $a(\tau)$, $b(\tau)$ and spatial coordinate $x=X(\tau)$
\begin{equation}\label{zhshArXiv:eq:6.14}
\frac{da}{d\tau}=-t_b(a,b), \quad
\frac{db}{d\tau}=t_a(a,b),
\end{equation}
\begin{equation}\label{zhshArXiv:eq:6.15}
\frac{dX}{d\tau}=(\lambda^2(r^1(b),r^2(a))-\lambda^1(r^1(b),r^2(a)))t_a(a,b) t_b(a,b),
\end{equation}
\begin{equation}\label{zhshArXiv:eq:6.16}
a\bigr|_{\tau=0}=a_*, \quad b\bigr|_{\tau=0}=b_*, \quad
X\bigr|_{\tau=0}=X_*.
\end{equation}
Integrating the problem (\ref{zhshArXiv:eq:6.14})--(\ref{zhshArXiv:eq:6.16}) we get the solution for the each parameter
$\tau$ on isochrone
\begin{equation}\label{zhshArXiv:eq:6.17}
R^1(x,t_*)=R^1_0(b(\tau)), \quad R^2(x,t_*)=R^2_0(a(\tau)), \quad
x=X(\tau).
\end{equation}
Moving along the isochrone we obtain solution for each values $x$ at fixed time $t=t_*$.
It is clear that the problem (\ref{zhshArXiv:eq:6.14})--(\ref{zhshArXiv:eq:6.16}) one should solve for $\tau>0$ and $\tau<0$.

We make a few remarks. The first, the equations (\ref{zhshArXiv:eq:6.14}) are only sufficient conditions for validity of equality (\ref{zhshArXiv:eq:6.12}). The right hand sides of differential equations (\ref{zhshArXiv:eq:6.14}) can be chosen accurately with arbitrary multiplier. That means that we can redefine the parameter $\tau$. In some cases, a good choice of the parameter $\tau$ allows to solve the Cauchy problem more effectively. The second, one \textbf{\emph{should not}} assume that  $\tau=x$. This is easily achieved by reduction of the  equations (\ref{zhshArXiv:eq:6.12}), (\ref{zhshArXiv:eq:6.13}) to equations
\begin{equation*}
\frac{da}{d\tau}=-\frac{1}{(\lambda^2-\lambda^1)t_a(a,b)}, \quad
\frac{db}{d\tau}=\frac{1}{(\lambda^2-\lambda^1)t_b(a,b)}, \quad
\frac{dX}{d\tau}=1.
\end{equation*}
At first sight, this replacement allows to reduce the number of equations and to get more natural form of the solution (\ref{zhshArXiv:eq:6.17}): $R^1(x,t_*)=R^1_0(b(x))$, $R^2(x,t_*)=R^2_0(a(x))$. However, this option does not allow to construct a multi-valued solution, in particular, it does not allow to study the breaking solutions.

In conclusion, we note that the right hand sides of equations (\ref{zhshArXiv:eq:6.14}), (\ref{zhshArXiv:eq:6.15}) are completely determined by the relations (\ref{zhshArXiv:eq:6.01})--(\ref{zhshArXiv:eq:6.05}) and, of course, by the  Riemann--Green function. Only the function $t(a,b)$ and its derivatives
are required for the calculations.


\setcounter{equation}{0}

\section{Classical shallow water equations}\label{zhshArXiv:sec:7.3}

To illustrate the effectiveness of the method we present the results of calculations for the shallow water equations.
The classic version of the shallow water equations without taking into account the incline of the bottom has the form (see \emph{i.g.} \cite{RozhdestvenskiiYanenko,Whithem})
\begin{equation}\label{zhshArXiv:eq:7.3C.01}
h_t+(hv)_x=0, \quad v_t+\left(\frac12v^2+h\right)_x=0,
\end{equation}
where $h>0$ is the elevation of the free surface, $v$ is the velocity.

We rewrite (\ref{zhshArXiv:eq:7.3C.01}) as
\begin{equation}\label{zhshArXiv:eq:7.3C.02}
u^1_t+(u^1 u^2)_x=0, \quad u^2_t+\left(\frac12 u^2 u^2+u^1\right)_x=0,
\end{equation}
\begin{equation}\label{zhshArXiv:eq:7.3C.03}
h=u^1, \quad v=u^2.
\end{equation}

The Riemann invariants for (\ref{zhshArXiv:eq:7.3C.02}) is well known
\begin{equation}\label{zhshArXiv:eq:7.3C.04}
R^1_t+\lambda^1 R^1_x=0, \quad R^2_t+\lambda^2 R^2_x=0,
\end{equation}
where
\begin{equation}\label{zhshArXiv:eq:7.3C.05}
\lambda^1(R^1,R^2)=\frac{3R^1+R^2}{4}=u^2-\sqrt{u^1},
\quad
\lambda^2(R^1,R^2)=\frac{3R^1+R^2}{4}=u^2-\sqrt{u^1},
\end{equation}
\begin{equation}\label{zhshArXiv:eq:7.3C.06}
u^1=\left(\frac{R^2-R^1}{4}\right)^2, \quad u^2=\frac{R^1+R^2}{2},
\end{equation}
\begin{equation}\label{zhshArXiv:eq:7.3C.07}
R^1=u^2-2\sqrt{u^1}, \quad R^2=u^2+2\sqrt{u^1}.
\end{equation}


\subsection{The function $\varphi(R^1,R^|r^1,r^2)$}\label{zhshArXiv:sec:7.3.1}

To determine the density $\varphi$ of the conservation law
\begin{equation}\label{zhshArXiv:eq:7.3C.08}
\varphi_t+\psi_x=0
\end{equation}
we use the Riemann--Green function for equation (\ref{zhshArXiv:eq:4.01})
\begin{equation}\label{zhshArXiv:eq:7.3C.09}
\Phi_{R^1 R^2} + A(R^1,R^2)\Phi_{R^1} + B(R^1,R^2)\Phi_{R^2}=0,
\end{equation}
\begin{equation*}
A(R^1,R^2)=\frac{\lambda^1_{R^2}}{\lambda^1-\lambda^2}=\frac{1}{2(R^1-R^2)}, \quad
B(R^1,R^2)=-\frac{\lambda^2_{R^1}}{\lambda^1-\lambda^2}=-\frac{1}{2(R^1-R^2)}.
\end{equation*}
The function $\Phi(R^1,R^2|r^1,r^2)$ is well known (see, \emph{i.g.} \cite{Copson}). Omitting the cumbersome transformations we only write the final result
for the density $\varphi(R^1,R^|r^1,r^2)$ (see, in particular, (\ref{zhshArXiv:eq:4.09}))
\begin{equation}\label{zhshArXiv:eq:7.3C.10}
\frac12\varphi=-\frac{2(R^1-R^2)^{1/2}}{(r^1-r^2)^{3/2}}
F
\left
(-\frac12,\frac32;1,-z
\right), \quad z=-\frac{(R^1-r^1)(R^2-r^2)}{(R^1-R^2)(r^1-r^2)},
\end{equation}
where $F$ is the hypergeometric function (see Appendix~\ref{zhshArXiv:sec:0A}).

We also write the derivatives of the function $\varphi$ with respect variables $r^1$, $r^2$, which are required for the calculation of the derivatives of $t_a$, $t_b$
\begin{equation}\label{zhshArXiv:eq:7.3C.13}
\frac12\varphi_{r^2}=\frac{3(R^1-R^2)^{1/2}}{(r^1-r^2)^{5/2}}
\left(
H_0(z)+\frac12 z_1 H_1(z)
\right),
\end{equation}
\begin{equation*}
\frac12\varphi_{r^1}=\frac{3(R^1-R^2)^{1/2}}{(r^1-r^2)^{5/2}}
\left(
-H_0(z)+\frac12 z_2 H_1(z)
\right),
\end{equation*}
\begin{equation*}
\quad z=-\frac{(R^1-r^1)(R^2-r^2)}{(R^1-R^2)(r^1-r^2)},
\end{equation*}
\begin{equation*}
\quad z_1=-\frac{(R^2-r^2)(R^1-r^2)}{(R^1-R^2)(r^1-r^2)}, \quad
\quad z_2= \frac{(R^1-r^1)(R^2-r^1)}{(R^1-R^2)(r^1-r^2)},
\end{equation*}
\begin{equation*}
H_0(z)=F\left(-\frac12,\frac32;1,-z\right)=F\left(\frac32,-\frac12;1,-z\right),
\end{equation*}
\begin{equation*}
H_1(z)=F\left(\frac12,\frac52;2,-z\right).
\end{equation*}

Using the results of Secs.~\ref{zhshArXiv:sec:2}--\ref{zhshArXiv:sec:6} we obtain the Cauchy problem for ODE's.


\subsection{Numerical results}\label{zhshArXiv:sec:7.3.2}


To demonstrate the effectiveness of the proposed method, we consider the evolution of the initial periodic free surface $h=u^1$ and periodic distribution of the velocity $v=u^2$
\begin{equation}\label{zhshArXiv:eq:7.3C.14}
u^1_0=1+0.1\cos x, \quad u^2_0=0.1\sin x.
\end{equation}

\begin{figure}[H]
\centering
\includegraphics[scale=1.00]{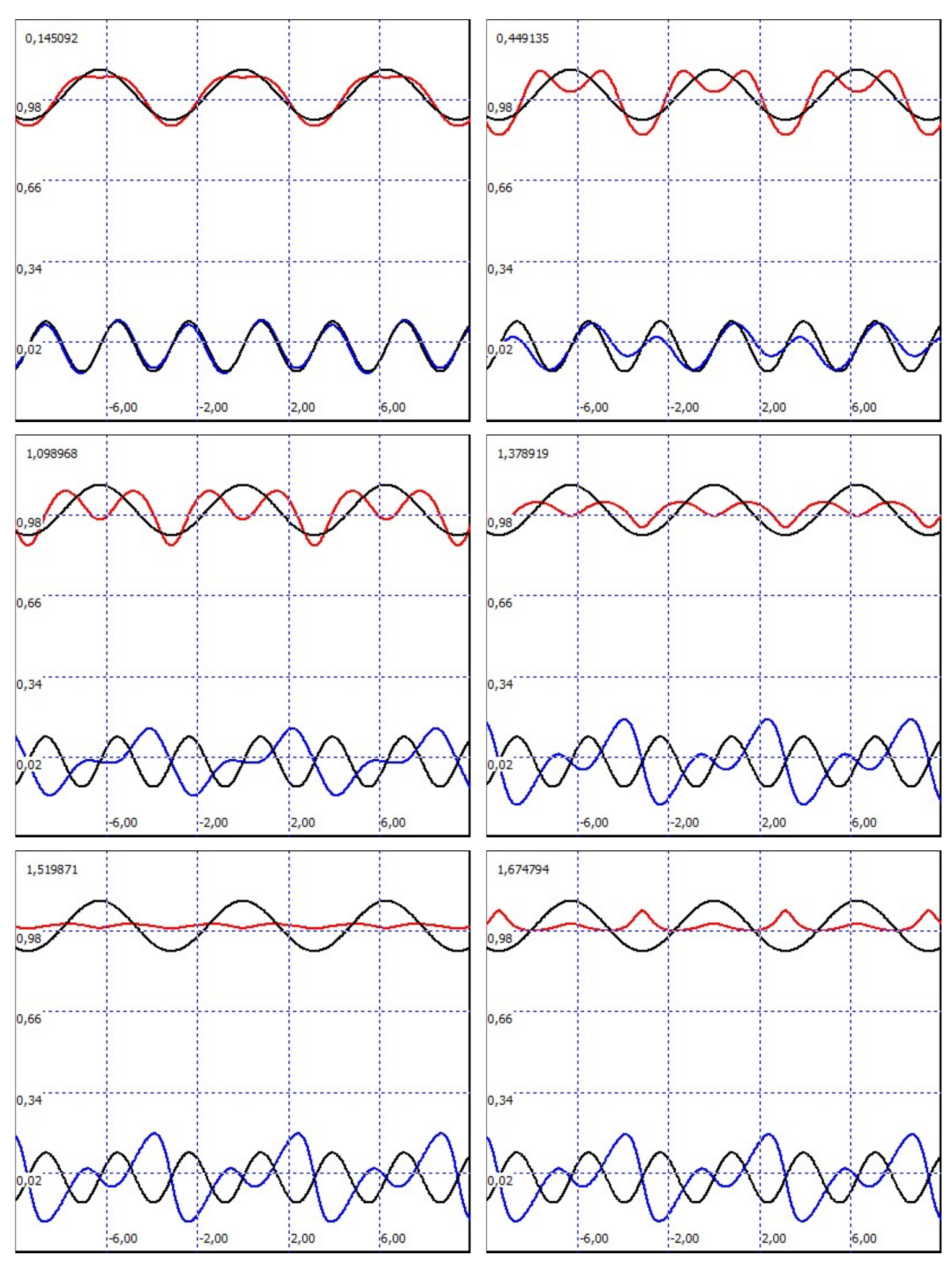}
\caption{$h=u^1$ (red line), $v=u^2$ (blue line), and initial distributions (black line), \\
$t=0.145$, $t=0.449$, $t=1.099$, $t=1.379$, $t=1.519$, $t=1.675$}
\label{fig13.2.3}
\end{figure}

\begin{figure}[H]
\centering
\includegraphics[scale=1.00]{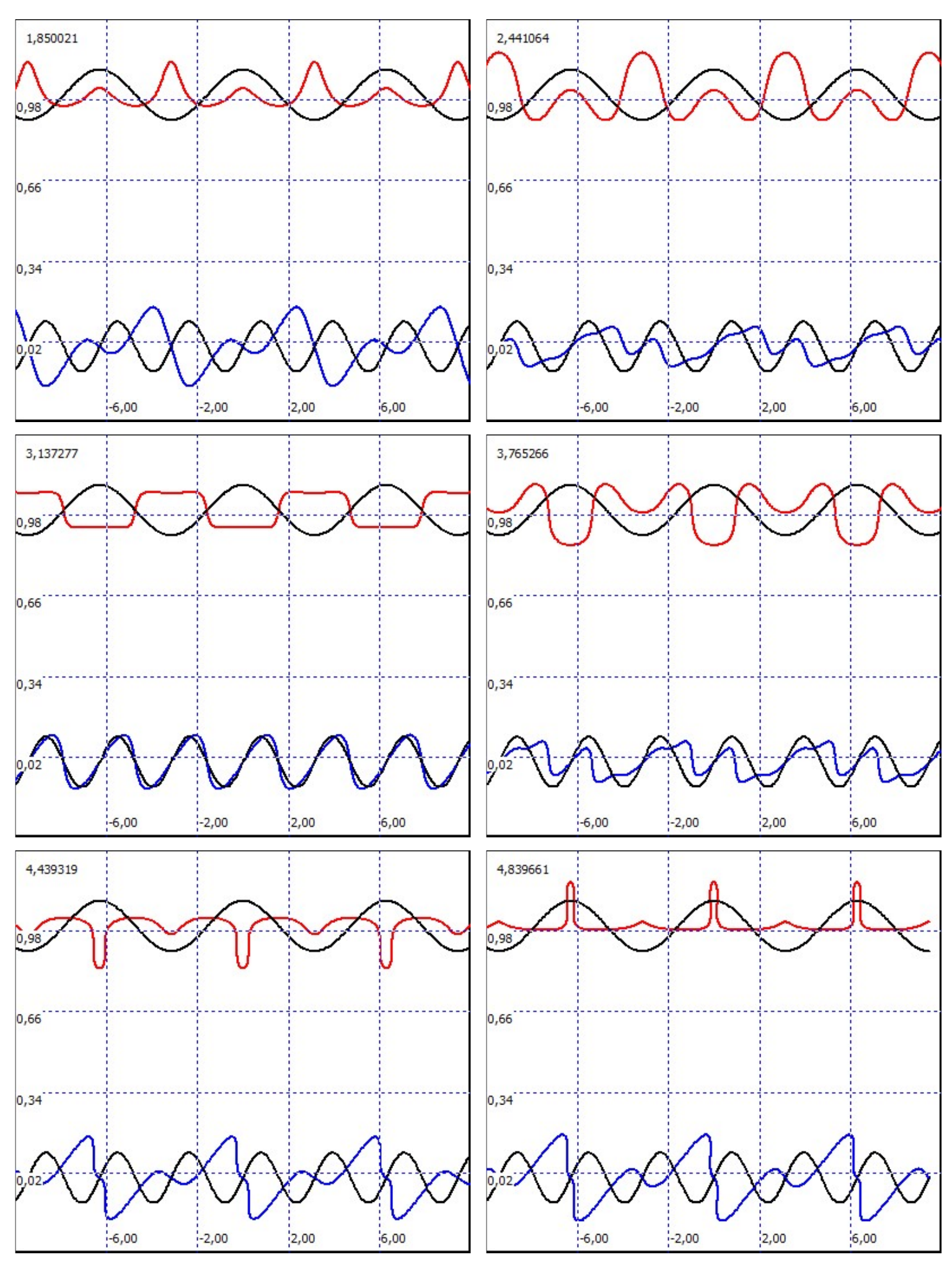}
\caption{$h=u^1$ (red line), $v=u^2$ (blue line), and initial distributions (black line), \\
$t=1.850$, $t=2.441$, $t=3.137$, $t=3.765$, $t=4.439$, $t=4.839$}
\label{fig13.2.6}
\end{figure}

\begin{figure}[H]
\centering
\includegraphics[scale=1.00]{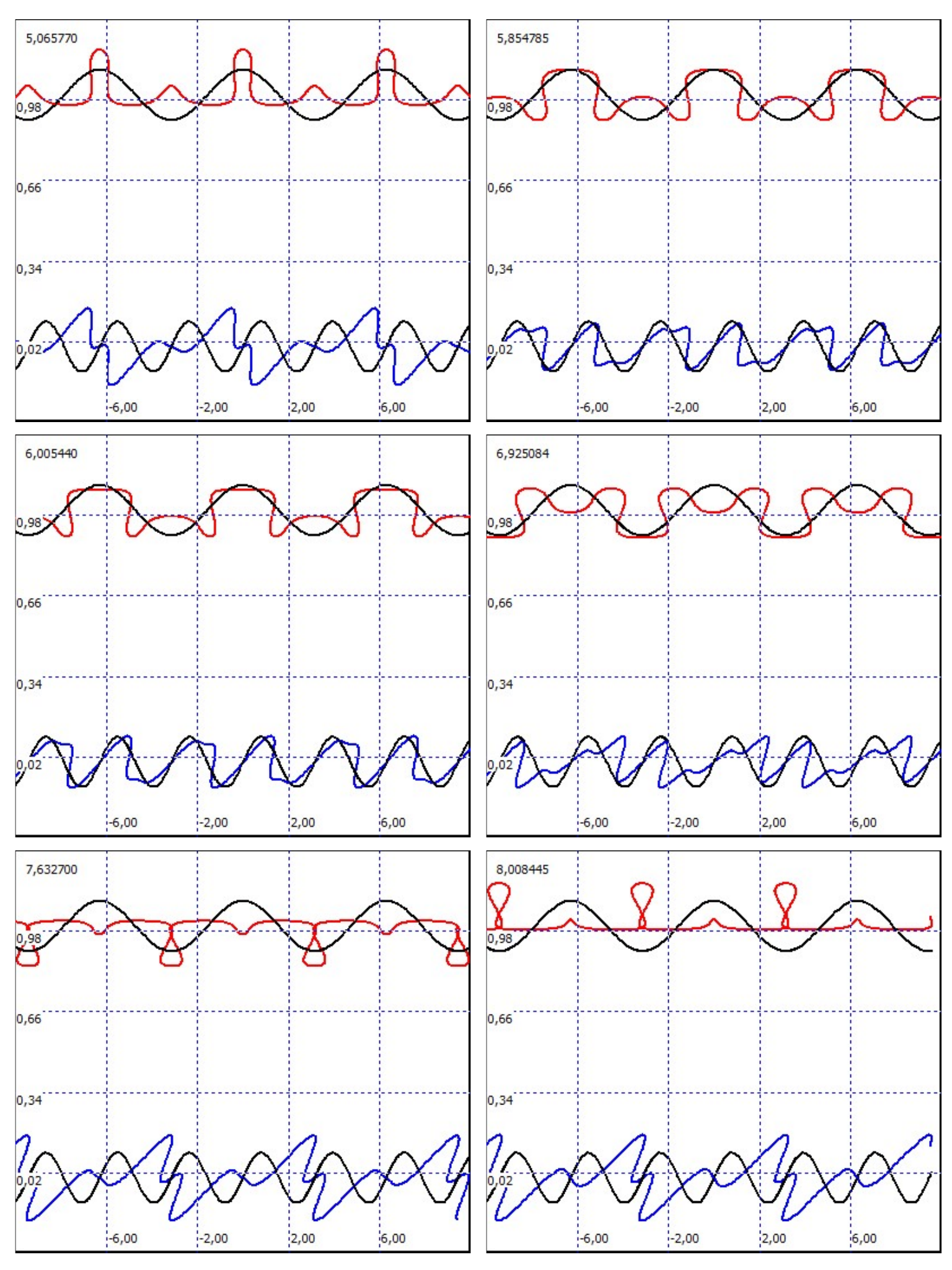}
\caption{$h=u^1$ (red line), $v=u^2$ (blue line), and initial distributions (black line), \\
$t=5.066$, $t=5.855$, $t=6.005$, $t=6.925$, $t=7.633$, $t=8.008$}
\label{fig13.2.10}
\end{figure}

\begin{figure}[H]
\centering
\includegraphics[scale=1.00]{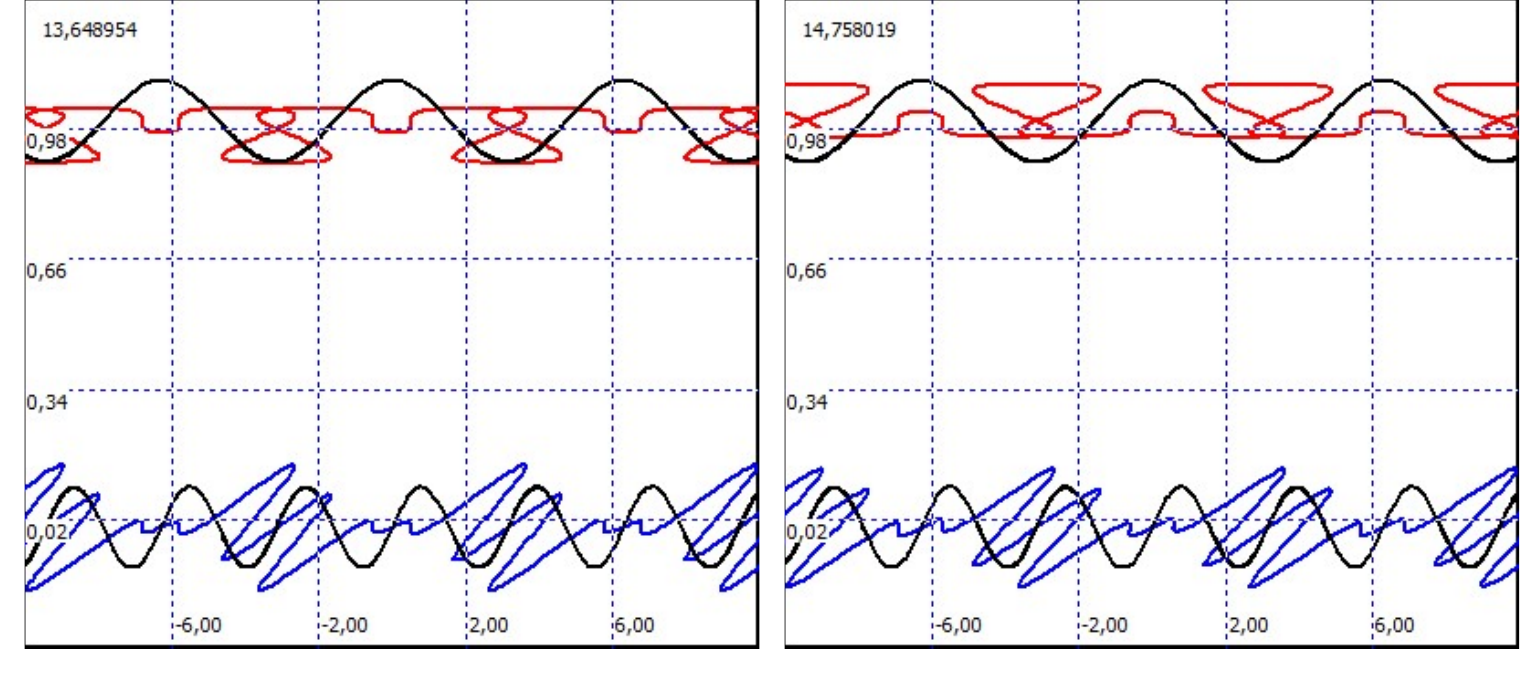}
\caption{$h=u^1$ (red line), $v=u^2$ (blue line), and initial distributions (black line), \\
$t=13.649$, $t=14.758$}
\label{fig13.2.11}
\end{figure}

The results of calculation  of the free surface position and the distribution of the velocity  for different time moments  are shown on Figs.~\ref{fig13.2.3}--\ref{fig13.2.11}.

The results obtained are not described in detail, and we only  restrict to a few comments.
On the Fig.~\ref{fig13.2.10} at $t=5.855$ the breaking waves clearly visible. Fig.~\ref{fig13.2.10} at $t=7.633$, $t=8.008$, and Fig.~\ref{fig13.2.11}  can be interpreted as the occurrence of bubbles or droplets on the shallow water surface.

\setcounter{equation}{0}

\section{Conclusions}

Implicit solution (\ref{zhshArXiv:eq:5.01}), (\ref{zhshArXiv:eq:5.02}) of the original problem (\ref{zhshArXiv:eq:2.01}), (\ref{zhshArXiv:eq:2.02}), of course, is very important to study of the original Cauchy problem properties. However, from our point of view, the implicit solution is not less complex than the original problem. For practical applications we must get, in one way or another, the explicit relations (\ref{zhshArXiv:eq:5.03}). In the general case we need to use a numerical methods, for example, Newton's method, for solving of the transcendental equations. This, in turn, requires a good initial approximations, or the using of the movement parameter method. It is especially difficult to numerically solve the system (\ref{zhshArXiv:eq:5.01}) in the case when original problem has multi-valued solitions. In other words, the using of the numerical methods for solving the systems of transcendental equations is not
much easier than the application of the direct methods for solving of the original problem, for example, using finite difference or finite volume method.
Instead solving of the systems of transcendental equations we propose to solve the Cauchy problem for ODE's. Even if it is need to solve ODE's numerically, for example, by the Runge--Kutta method, the numerical algorithm is realized simpler than the algorithm for solving the system of nonlinear transcendental equations.

In the next papers we plan to present the results of the calculations for equations of the zonal electrophoresis and  the soliton gas equations.

\begin{acknowledgments}
The authors are grateful to N. M. Zhukova for proofreading the manuscript.
Funding statement. This research is partially supported by the Base Part of the Project no. 213.01-11/2014-1,
Ministry of Education and Science of the Russian Federation, Southern Federal University.
\end{acknowledgments}

\appendix

\section*{Appendix}\label{App:ZhS-0}

\renewcommand{\theequation}{A\arabic{section}.\arabic{equation}}%
\setcounter{equation}{0}

\section{Hypergeometric function and Elliptic integrals}\label{zhshArXiv:sec:0A}

For practical calculations of the hypergeometric function (see, (\ref{zhshArXiv:eq:7.3C.10}), (\ref{zhshArXiv:eq:7.3C.13})) one can use the
complete elliptic integrals $\textrm{\textbf{E}}$, $\textrm{\textbf{K}}$.
\begin{equation}\label{zhshArXiv:eq:7.3C.11}
F\left(-\frac{1}{2},\frac{3}{2};1,-z\right)=F\left(\frac{3}{2},-\frac{1}{2};1,-z\right)=
\end{equation}
\begin{equation*}
=-\frac{2}{\pi\sqrt{1+z}} \textrm{\textbf{K}}
\left(\frac{\sqrt{z}}{\sqrt{1+z}}\right)
+ \frac{4\sqrt{1+z}}{\pi} \textrm{\textbf{E}}
\left(\frac{\sqrt{z}}{\sqrt{1+z}}\right),
\end{equation*}
\begin{equation*}
H_0(z)=F\left(-\frac12,\frac32;1,-z\right)=F\left(\frac32,-\frac12;1,-z\right)=
\end{equation*}
\begin{equation*}
=-\frac{2}{\pi\sqrt{1+z}} \textrm{\textbf{K}}
\left(\frac{\sqrt{z}}{\sqrt{1+z}}\right)
+ \frac{4\sqrt{1+z}}{\pi} \textrm{\textbf{E}}
\left(\frac{\sqrt{z}}{\sqrt{1+z}}\right),
\end{equation*}
\begin{equation*}
H_1(z)=F\left(\frac12,\frac52;2,-z\right)=
\end{equation*}
\begin{equation*}
=\frac{1}{3\pi z \sqrt{1+z}}
\left(
-4\textrm{\textbf{K}}
\left(\frac{\sqrt{z}}{\sqrt{1+z}}\right)
+ (4+8z)\textrm{\textbf{E}}
\left(\frac{\sqrt{z}}{\sqrt{1+z}}\right)
\right),
\end{equation*}
\begin{equation*}
\lim_{z\,\to\, +0} H_1(z)=1.
\end{equation*}

Pay attention, that when we compute the complete elliptic integrals of $\textrm{\textbf{E}}$, $\textrm{\textbf{K}}$
on the interval $-1<z<0$ then the arguments of the functions $\textrm{\textbf{E}}$, $\textrm{\textbf{K}}$ are imaginary. In the absence of complex arithmetic for calculations one can use the relations
\begin{equation}\label{zhshArXiv:eq:7.3C.12}
\textrm{\textbf{K}}
\left(\frac{\sqrt{z}}{\sqrt{1+z}}\right)
=\sqrt{1+z}\textrm{\textbf{K}}(\sqrt{|z|}), \quad -1 < z < 0,
\end{equation}
\begin{equation*}
\textrm{\textbf{E}}
\left(\frac{\sqrt{z}}{\sqrt{1+z}}\right)
=\frac{1}{\sqrt{1+z}}\textrm{\textbf{E}}(\sqrt{|z|}), \quad -1 < z < 0.
\end{equation*}

\setlength{\bibsep}{4.0pt}


\begin{thebibliography}{99}
\small

\bibitem{SenashovYakhno}
{\it Senashov~S.\,I., Yakhno~A.\/} Conservation laws, hodograph transformation and boundary value problems of plane plasticity. 2012. SIGMA.   Vol.\,8, 071. 

\bibitem{RozhdestvenskiiYanenko}
{\it Rozdestvenskii B.L., Janenko N.N.} Systems of Quasilinear Equations and Their Applications to Gas
Dynamics [in Russian], Nauka, Moscow (1978); English transl.: Transl. Math. Monogr., Vol. 55, Amer. Math.
Soc., Providence, R. I. (1983).


\bibitem{Whithem}
{G.B.Whithem, Linear and nonlinear wave. A Wiley-Interscience Publication John Willey \&
Sons, 1974, New-York--London--Sydney--Toronto.}

\bibitem{GenaEl}
\textsl{El. G.A., Kamchatnov A.M.}  Kinetic equation for a dense soliton gas. 2006. ArXiv:nlin/0507016v2.  

\bibitem{FerapontovTsarev_MatModel}
{\it Ferapontov E. V., Tsarev S. P.} Ferapontov, E. V.; Tsarev, S. P. Systems of hydrodynamic type that arise in gas chromatography. Riemann invariants and exact solutions. 1991.  Math. Model. 3 (1991), no.~2, 82–-91. (Russian)


\bibitem{Kuznetsov}
{\it Kuznetsov N.\,N.\/}  	
Some mathematical questions of chromatography.  Computation methohods and programming. 1967.  no.\,6, 242--258.


\bibitem{ElaevaMM}
{\it Elaeva M. S. \/} Investigation of zonal elecrophoresis for two component mixture. 2010.  Math. Model. 22, no. 9, 146–-160. (Russian)

\bibitem{Elaeva_ZhVM}
{\it Elaeva M. S.\/} Separation of two component mixture under action an electric field. 2012.  Comp. Math. and Mat. Phys. 52:6, 1143–-1159.

\bibitem{BabskiiZhukovYudovichRussian}
{\it Babskii V. G., Zhukov M. Yu., Yudovich V. I.\/} Mathematical Theory of Electrophoresis.
Kluwer Academic / Plenum Publishers (1989).


\bibitem{ZhukovNonSteadyITP}
{\it Zhukov M. Yu. } Non-staionary isotachophoresis model. 1984. Comp. Math. and Math. Phys. Vol.\,24, No\,4, 549--565. (in Rissian).


\bibitem{ZhukovMassTransport}
{\it Zhukov M. Yu.\/} Masstransport by an electric field. Rostov-on-Don: RGU Press, 2005. 


\bibitem{Copson}
{\it Copson E. T.\/} On the Riemann-Green Function.Arch. 1958. Ration. Mech. Anal. 1,  324--348.

\bibitem{Courant}
\textsl{Courant R., Hilbert D.} Methods of Mathematical Physics: Partial Differential Equations, Volume II.  New York -- London, 1964.



\bibitem{Ibragimov}
{\it Ibragimov~N.\,Kh.\/} Group analysis of ordinary differential equations and the invariance principle in mathematical physics (for the 150th anniversary of Sophus Lie). 1992.  Russian Mathematical Surveys, 47(4):89. 83--144.


\bibitem{Chirkunov}
\textsl{Yu. A. Chirkunov} On the symmetry classification and conservation laws for quasilinear differential equations of second order. 2010. Mathematical Notes.  Vol. 87, (1-2). 115--121.


\bibitem{Chirkunov_2}
\textsl{Yu. A. Chirkunov}  Generalized equivalence transformations and group classification of systems of differential equations. 2012. Journal of Applied Mechanics and Technical Physics. Vol. 53 (2). 147--155.


\end{thebibliography}
\end{document}